\documentclass{revtex4}
\usepackage{graphicx}

\begin{document}

\title{Nanoscale buckling deformation in layered copolymer materials}


\author{Ali Makke $^1,^2$, Michel Perez$^2$, Olivier Lame$^2$, Jean-Louis Barrat$^3$}
\affiliation{
$^1$  Universit\'e de Lyon- Univ. Lyon I - LPMCN - UMR CNRS 5586- F69622 Villeurbanne, France\\
$^{2}$ Universit\'e de Lyon - INSA Lyon - MATEIS - UMR CNRS 5510 - F69621 Villeurbanne, France\\
$^{3}$Univ. Grenoble 1 / CNRS, LIPhy UMR 5588, Grenoble, F-38041, France}


\begin{abstract}
In layered materials, a common mode of deformation 
involves buckling of the layers under tensile deformation in the direction perpendicular to the layers.  
The instability mechanism, 
which operates  in elastic materials from geological to nanometer scales, 
involves the elastic contrast 
between different layers. In a  regular stacking of ``hard'' and ``soft'' layers, 
the tensile stress is first accommodated by a large deformation of the soft layers. The inhibited Poisson contraction results in a 
 compressive  stress  in the direction transverse to 
the tensile deformation axis. The  ``hard''  layers sustain this transverse compression until buckling takes place and results in 
an undulated structure. Using molecular simulations, we  demonstrate this scenario  for a material made of triblock copolymers. 
The buckling deformation is observed to take place at the nanoscale, at a wavelength that depends on strain rate.
In contrast to what is commonly assumed, the wavelength of the undulation is not determined by  defects in the microstructure. 
Rather, it results from kinetic effects, with a competition between the rate of strain and the growth rate of
the instability.
\end{abstract}
\maketitle



The mechanical response of multiblock copolymers depends sensitively upon their constituent homopolymers segments, molecular architecture and chain topology.
Triblock copolymers, in particular, have become an attractive material for their use as  thermoplastic elastomers that could be  integrated in 
several technical and manufactural fields ( copolymer styrene butadiene rubber is commercially exploited in footwear, in pressure sensitive adhesive (K-Resine), in paving and roofing compounds....).
Depending on the amount of each phase the segregated block copolymers may present  several morphologies e.g. spherical, cylindrical, lamellar... \cite{Bates99}. 
The lamellar morphologies are particularly interesting as model systems, as well aligned specimens can be prepared by shearing. The one dimensional aspect of the structure simplifies the analysis, and the mechanical response reveals equally the presence of both phases, while in other morphologies
it tends to be dominated by the majority matrix phase.
In these block  copolymer systems the constituent blocks are generally chosen such that one of them is glassy and the other one is rubbery. 
 A single copolymer chain can be shared between two different glassy lamellae,  forming  a rubber bridge that provides a strong coupling between phases. 
 The resulting system combines the stiffness of the hard glassy phase and the ductility of the soft rubbery phase.

\begin{figure}[tbp]
 \centering\includegraphics[width=0.4\textwidth]{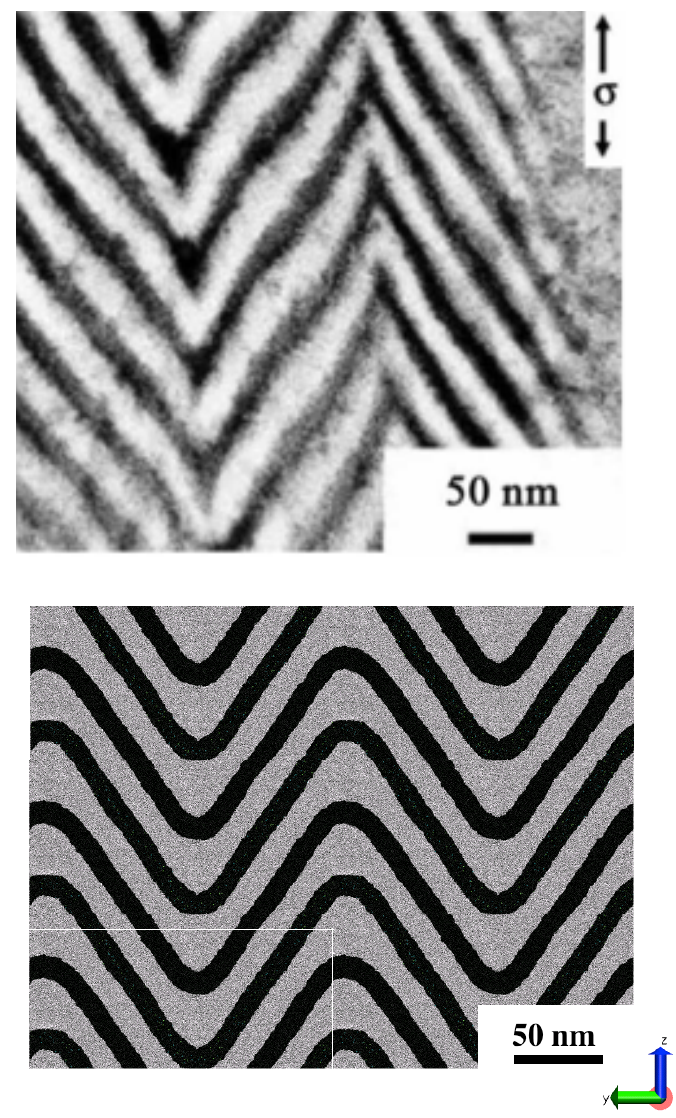}
\caption{\label{sample_real.fig}
Top panel: Transmission electron micrograph shows the buckling observed in a real copolymer sample (Styrene-Butadiene-styrene). The figure is extracted from reference \cite{Adhikari04}. Bottom panel: Lamellae buckling observed by molecular dynamics simulation, snapshot is taken at $e_{zz}=0.55$. This image contains six periodic copies of the simulated system, which is enclosed in the white frame. The scale on the right is obtained by assuming a monomer size $\sigma=0.5nm$.}
\end{figure}

When such  a lamellar copolymer sample is submitted to a tensile strain perpendicular to the layers, the glassy layer eventually buckles into  a ``chevron'' morphology.
With increasing strain the  normal to the lamellae tilts away from the stretching direction, whereas the lamellar  spacing remains almost constant. This behavior
 was demonstrated experimentally in triblock copolymers  by Small Angle Xray Scattering (SAXS) under deformation \cite{Cohen_1_01}, and by micrographs of strongly 
 deformed samples \cite{Cohen_2_01} (see figure \ref{sample_real.fig}). Similar effects are also observed in lamellar systems with alternating crystalline and glassy parts \cite{Hermel_03,Phatak_06}, and in deformed semicrystalline polymers, which form locally 
 lamellar structures of crystalline material separated by softer amorphous parts \cite{Krumova06}.

This buckling instability under strain,   which is observed in many layered materials from smectic liquid crystals \cite{Gennes93} to geological layers  \cite{Biot61,Ramberg64},
 is frequently described in a qualitative way by a preference to shear compared  to an extension in the direction normal  to the layers, in order to 
 preserve the lamellar spacing. 
 A different cause for buckling is the existence of a Poisson effect. In a stacking of ``hard'' and ``soft'' layers, 
the soft  phase accommodates  most of the imposed deformation, and as the Poisson contraction is prevented by the 
coupling to the hard phase; it  exerts a compressive stress in the  transverse direction. 
  The``hard'' layers sustain this transverse
compression until buckling takes place and results in an undulated structure. In general, elasticity
predicts buckling to take place on the largest wavelength compatible with the boundary conditions
imposed to the system \cite{Read99}. We study this generic scenario by means of molecular dynamics simulations,
for a material made of triblock copolymers in their lamellar phase. The contrast in elasticity is provided by a different glass transition temperature of the different blocks.  We  use the ability of molecular dynamics simulations  to give information on the local values of stresses and strains to explore the causes of the instability in triblock copolymers 
 with alternating glassy and rubbery layers,  without introducing an a priori description of the mechanism. While the elastic origin of the instability is confirmed, 
 our results demonstrate an unexpected dependence of the failure mode on strain rate. The  wavelength of the undulations is determined by the deformation rate, and a characteristic size emerges  even in the absence of preexisting defects in the microstructure.  We discuss a general mechanism for the emergence of 
 such a characteristic size, which results from a competition between the rate of strain
and the growth rate of the buckling instability.

\section{Evidence of the Poisson effect}
\label{sec:poisson}
When a multilayered system is stretched perpendicularly to the layering direction, each component of the system will deform according to its own
stiffness. Locally, the deformation is distributed between phases in a way that ensures the continuity of the  stress. The resulting macroscale 
deformation is the sum of the local strain response of each phase, so that  the  stiffness will be dominated by the response of the soft phase.
In a lamellar copolymer with alternating  rubbery and glassy phases,  the tensile strain will be mainly localized in the rubbery lamellae, which are essentially incompressible (Poisson ratio  $\nu_{rubbery} \simeq 0.5$). The contraction in the transverse direction that would result from the Poisson effect   is however  prevented by the strong coupling  to the hard phase at the interface, so that the soft phase exerts a strong compressive stress  in the transverse direction
on the hard layers.
The glassy phase becomes submitted to a tensile stress in the perpendicular direction and a compressive stress laterally. 
Under these conditions,  and for a sufficiently large system,   the buckling instability takes place to relax the lateral compressive stress.

\begin{figure}[tbp]
\includegraphics[width=0.48\textwidth]{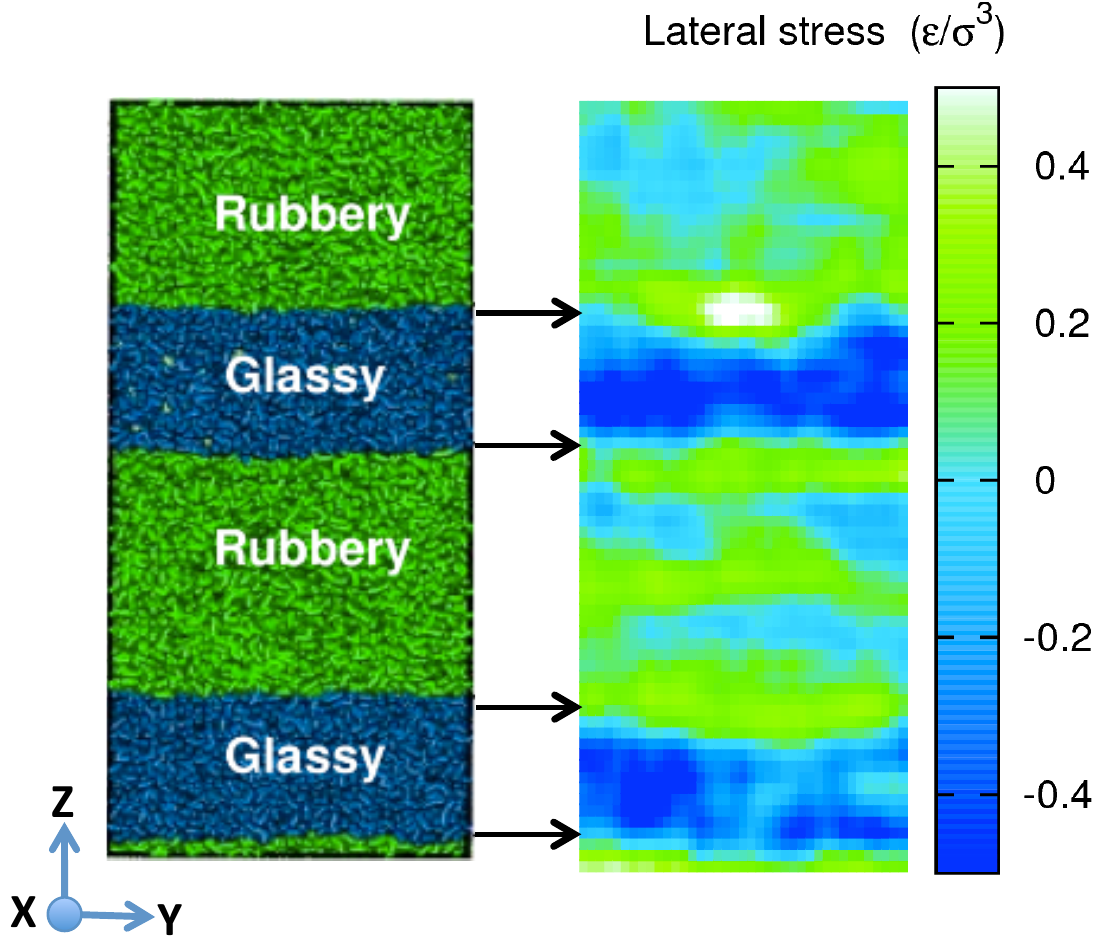}
\caption{\label{profile_stress_per_phase.fig}
Cartography of the lateral stress $\frac{\Sigma_{xx}+\Sigma_{yy}}{2}$. 
Snapshots were taken at a deformation $e_{zz}=0.04$.
The local stress is computed within a box of size $5\sigma\times5\sigma\times L_y=102 \sigma$, where $\sigma$ is the
 size of a monomer (typically $\sigma=0.5nm$, see section \ref{sec:methods}). 
The stress in these figures was probed in a relatively small sample ($L_x=34.5\sigma$, $L_y=102.3\sigma$ and $L_z=34.5\sigma$).
 in which  buckling  does not take place, however the analysis of the stress profile clearly shows the importance of the Poisson effect, with a large contrast in the transverse stress between the compressed  glassy layers and the rubbery layers under tension . 
 The influence of the lateral dimension of the sample is discussed below.}
\end{figure}

This scenario can be checked by monitoring the local stress in a sample strained perpendicularly to the lamellae.
Figure \ref {profile_stress_per_phase.fig} shows a map of the lateral stress ($\frac{\Sigma_{xx}+\Sigma_{yy}}{2}$)
 of a stretched copolymer sample at a true strain $e_{zz}=0.04$. The average pressure in the transverse direction is   zero,
  as uniaxial tensile conditions are imposed globally.   The local  stress, however,  is positive in the rubbery phase while it is negative in the glassy phase,
   indicating a local compression parallel to the glassy lamellae. An interesting additional feature that is apparent in figure \ref{profile_stress_per_phase.fig}   
   is the existence of noticeable fluctuations in the local stress values. Such fluctuations are expected in the response of amorphous materials, and arise from the local heterogeneity of elastic properties in such materials,  a property already well documented for glassy polymers \cite{Yoshimoto_04}. We will see below that the response of the system can become macroscopically non uniform due to the development of an elastic instability, and it is therefore quite natural to speculate that the initial stress fluctuations serve to nucleate this instability. A related study of cavitation in glassy homopolymers showed a similar correlation between elastic fluctuations and nucleation of cavities \cite{Makke_11}.
   
\section{Plastic deformation through buckling and cavitation}
\label{sec:plastic}

\begin{figure}
\setlength\unitlength{1cm}
\includegraphics[width=0.48\textwidth]{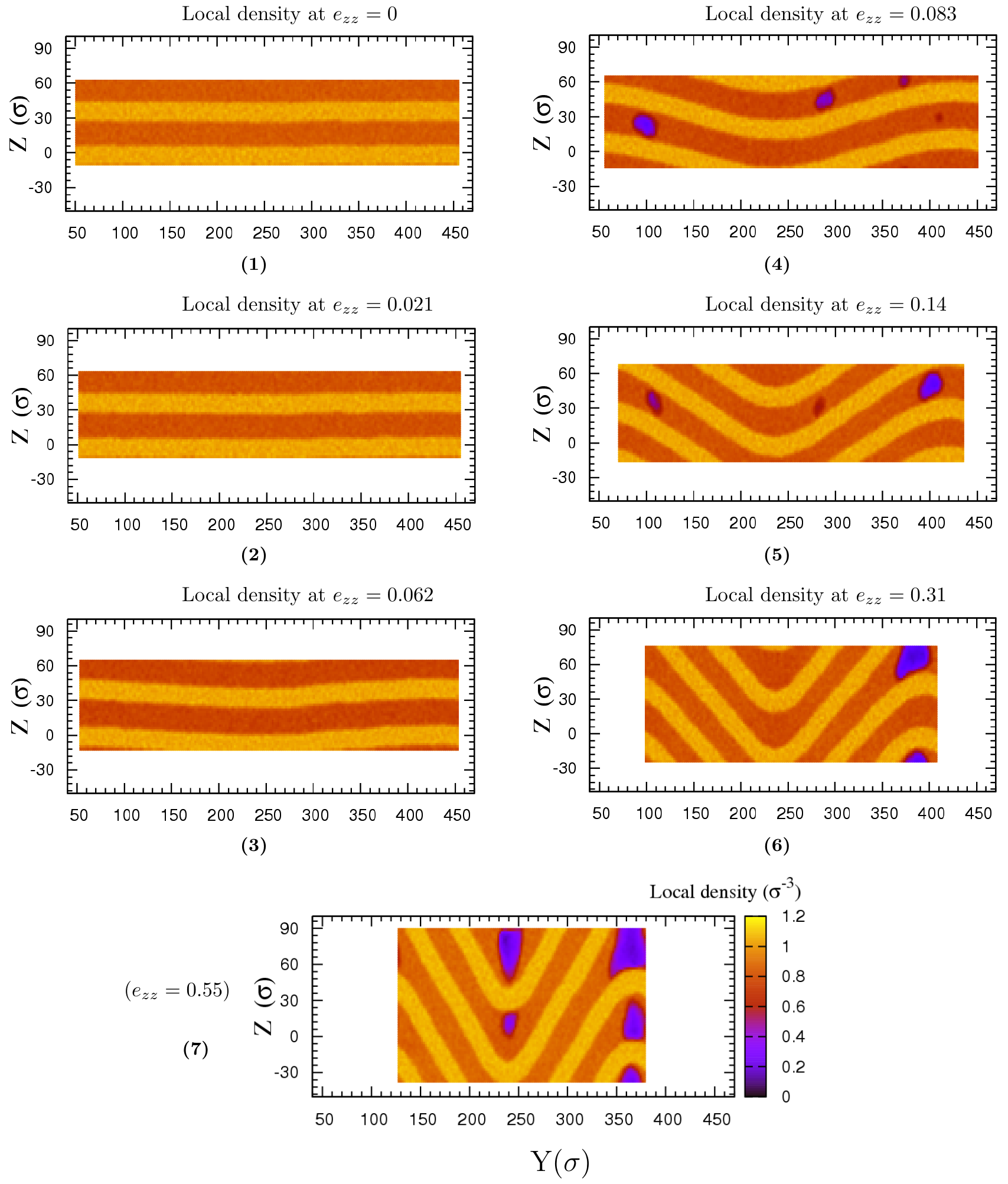}
\caption{\label{local_density}
Local density cartographies at several strains. The high (low) density phase corresponds to the glassy (rubbery) lamellae. The phase buckling starts before cavitation. 
As the deformation progresses, cavities will nucleate randomly in the rubbery phase. Cavities that are located in the tilted part of the chevron   disappear rapidly; however, only cavities that are located in the hinges will survive to a  high strain. }
\end{figure}

\begin{figure}
 \centering
\includegraphics[width=0.4\textwidth]{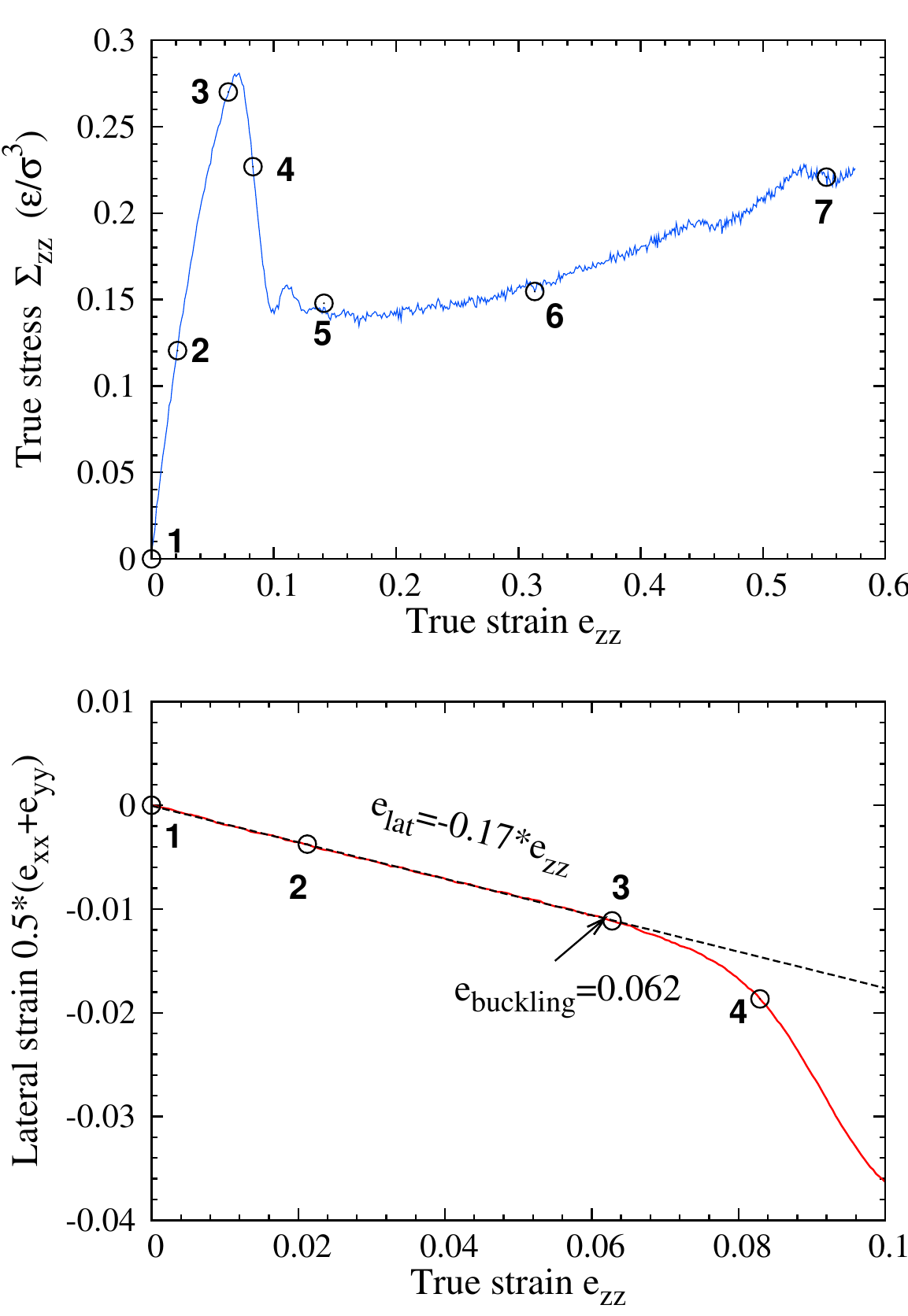}
\caption{\label{behavior_curve.fig}Top panel: stress strain curve  of the sample shown in figure \ref{local_density} under uniaxial strain conditions.
 The black points denote the selected configurations shown in figure \ref{local_density}. 
 The stress release is associated with a combination of two phenomena: buckling, and nucleation and growth of  cavities in the rubbery phases. Bottom panel:  evolution of the lateral strain $0.5(e_{xx}+e_{zz})$ under uniaxial tensile conditions.  The local slope defines the Poisson ratio, and buckling is correlated to a strong change in this slope, as the sample starts deforming in an accordion-like manner. }
\end{figure}

Figure  \ref{local_density} illustrates the deformation process that takes place under 
tensile deformation in a sufficiently large sample, and the corresponding stress-strain curve is shown in figure \ref{behavior_curve.fig}.
 In this first 
set of results, the strain rate is $\dot{e}_{zz} = 7.3\times10^{-5}$ in reduced units (see Materials and Methods section) . 
The pressure is fixed to  zero in the $X$ and $Y$ directions.

The stress strain curve displays three main regimes:
\begin{itemize}
\item
\textbf{Elastic regime: }  at small strain, the stress grows linearly with the imposed deformation. This regime is limited to a very small deformation amount (less than 2\%). The Young modulus, which can be fitted from this curve, results from   a combination of the elastic behavior of each phase.
The density maps show very little change in this regime.

\item
\textbf{Buckling :} Beyond the elastic regime, a progressive softening is observed. This slight deviation from the elastic linear behavior is commonly interpreted by the change of molecular conformation  in the rubbery phase. The buckling of the glassy phase starts at $e_{zz}= 0.06$ and  can be detected from a rapid change 
in the apparent Poisson ratio of the material. From the onset of the buckling instability, the sample starts deforming in an ``accordion'' like manner, with a higher Poisson ratio  than a sample with parallel layers.

\item                                                                                                                                                                                                                                                                                                                                                                                                                                                                                                                                                                                                                      
\textbf{Cavitation: }
After  buckling, a strong stress drop occurs at $e_{zz}= 0.08$. This drop can be  correlated with the nucleation of cavities in the rubbery phase, as illustrated by the fourth density map in figure \ref{local_density}. The low density spots in figure \ref{local_density} correspond to the cavitation in the rubbery layers.
Indeed, due to the buckling, the local deformation of the rubbery phase is not homogenous. The sample progressively adopts a chevron 
morphology, with different deformation states: at large strains the cavitation is confined to
a localized region in space, developing into a hinge.
 At the hinges of the chevron the deformation is essentially tensile, while
the tilted  part undergoes a simple shear deformation.
The latter  is caused by the rotation and sliding of the hard lamella.  In contrast, 
at  the hinges of the chevron the deformation  is essentially triaxial, and favors nucleation of cavities. 
As a result the cavities in the rubber that initially appear randomly tend to  heal in the sheared zones and nucleate preferentially where triaxial stress persists, as illustrated by the sequence of snapshots  in figure \ref{local_density}. ).
\end{itemize}

\section{Elastic description of the buckling instability}
\label{theory.sec}

As mentioned in the introduction, two different theoretical descriptions of buckling of lamellae under stress
are available in the literature. One approach is based on  a free energy functional of the order parameter 
that describes the lamellar order, and explains the instability by the fact that a strained state will try to maintain the
lamellar distance  that minimizes this free energy \cite{Wang93}. This approach would be appropriate for copolymers in which both phases are at equilibrium, 
so that the order parameter can respond to the deformation. Here one of the phases is glassy, and the explanation for the
buckling instability must be searched in a different direction, involving the minimization of the total elastic energy. This 
description (which for some aspects goes back to early works of Biot \cite{Biot61,Biot63,Biot64}) has been detailed in a seminal paper by 
 Read \textit{et al} \cite{Read99}.  Taking for simplicity a two dimensional geometry, 
the elastic energy of the  sample under small strain  can be written as:
\begin{equation}
\label{ela_ene.eq}
U_{macro}=\frac{1}{2}(C_{11}e_{11}^2+2C_{12}e_{11}e_{22}+C_{22}e_{22}^2+Ge_{12}^2)
\end{equation}
where $C_{ij}$ are the components of a symmetric $2 \times 2$ stiffness matrix of \textit{the entire system}, and $e_{ij}$ is the macroscopic deformation.  
In this  equation the material is considered as homogeneous and  anisotropic. The indexes $1,2$ refer to the directions normal and parallel to the layers, and the coefficients $C_{ij}$ and $G$ (shear modulus parallel to the layers)    can be  obtained from mechanical tests in different directions, or deduced from the elastic properties of each lamella.

The  bending energy results from the variation of the lamellar rotation angle $\theta$ (the angle between the layer and the horizontal axis $X$ ).
The associated energy density reads
\begin{equation}
U_{bend}=\frac{1}{2}K(\nabla_x\theta)^2
\label{bend_ene.eq}
\end{equation}
 where $K$ is the bending modulus of the sample. Due to the serial coupling between phases the bending 
 modulus will be dominated by the contribution of the  hard phase.  The bending modulus can  then be estimated from simple beam bending theory as 
$K_{est}=\frac{\phi_h^3 E_h d^2}{12(1-\nu_h^2)}$ where $\phi_h$ is the volume fraction of the hard phase, $E_h$ is the Young modulus of the hard phase,
 $\nu_h$ is its Poisson ratio and $d$ is the lamellar spacing.

The total energy density results from the addition of the bulk elastic energy and the bending energy, $U_{2D}= U_{bend} +U_{macro}$.
It can be expressed in terms of the global deformation 
 $e_{xx}$, $e_{xz}$ and $e_{zz}$  and of a local, non affine displacement field $\vec{u}(x,z)$ that quantifies the difference between the displacement imposed at the global scale and the one observed locally.

 The final expression for the free energy, written as an expansion in powers of $\vec{u}(x,z)$ \cite{Read99} reads 
   \begin{equation}
\begin{array}{l}
2\langle U_{2d}\rangle
=C_{11}e_{xx}^2 + 2C_{13}e_{xx}e_{zz}+C_{33}e_{zz}^2 \\
\\
+  C_{11}\langle (\nabla_x u_x)^2 \rangle+  2C_{13}\langle (\nabla_z u_x)(\nabla_x u_z) \rangle+ 
 C_{33}\langle (\nabla_z u_z)^2 \rangle  \\
\\
 +\frac{\langle (\nabla_x u_z)^2 \rangle}{(1+e_{xx})^2} [G-e_{zz}( C_{33}- C_{13}-2G)- e_{zz}^2( C_{33}-G)\\
 \\
  +e_{xx}( C_{11}(1+e_{xx})- C_{13})] + K\langle (\nabla_x^2 u_z)^2 \rangle \\
  \\
+2\frac{\langle (\nabla_z u_x)(\nabla_x u_z) \rangle}{(1+e_{xx})}[G-e_{zz}(C_{33}- G)-e_{xx}C_{13}] 
+O(u^4)

\end{array}
\label{exp_energy.eq}
\end{equation}
A linear stability analysis of this energy is performed by introducing a  sinusoidal perturbation
of the form observed in our simulations, 
\begin{equation}
u_z(x,z) =U_0 \sin(kx)\ \ ; \ \ u_x(x,z)=0
\label{uz.eq}
\end{equation}
 with $k=2n\pi/L$  a  wave vector compatible with the boundary conditions,
 \begin{equation}
\begin{array}{l}
2\langle U_{2d}\rangle =C_{11}e_{xx}^2 + 2C_{13}e_{xx}e_{zz}+C_{33}e_{zz}^2 \\
\\
+\frac{1}{4} \{ f_1(e_{xx},e_{zz})k^2+Kk^4\} U_0^2+O(U_0^4)

\end{array}
\label{exp_energy_2.eq}
\end{equation}
where
\begin{equation}
\begin{array}{l}
f_1^{2D}(e_{xx},e_{zz}) = \frac{1}{(1+e_{xx})^2}[ G-e_{zz}( C_{33}- C_{13}-2G) \\
\\
- e_{zz}^2( C_{33}-G) +e_{xx}( C_{11}(1+e_{xx})- C_{13})]
\end{array}
\label{f1_2D.eq}
\end{equation}
The buckling instability occurs upon increasing strain when the coefficient of $U_0^2$  in equation \ref{exp_energy_2.eq}
becomes negative, meaning that the global gain in elastic energy 
 overwhelms the bending energy penalty. 
 In 3D case the same analysis process can be performed, this leads to a function $f_1^{3D}(e_{xx},e_{yy},e_{zz})$ instead $f_1^{2D}(e_{xx},e_{zz})$ where:

 \begin{equation}
\begin{array}{l}
f_1^{3D}(e_{xx},e_{zz},e_{yy})=f_1^{2D}(e_{xx},e_{zz}) - \frac{1}{(1+e_{xx})^2}  C_{23} e_{yy}
\end{array}
\label{f1_3D.eq}
\end{equation}
  
 To close the system, one assumes that before the  buckling begins (i.e.  in the elastic regime) $e_{xx}$ and $e_{yy}$ can be substituted
  by $\nu e_{zz}$ where $\nu$ is a global Poisson ratio.  Under this assumption, $f_1^{3D}(e_{xx},e_{yy},e_{zz})$ becomes a function $f_2^{3D}(e_{zz})$
 of $e_{zz}$ only. 
The buckling strain $e_{buck}^*$ can be estimated by solving this equation for a fixed wavevector $k_n = 2n\pi/L$:
\begin{equation}
 f_2^{3D}(e_{buck}^*)k_n^2+Kk_n^4 = 0
\label{buck_starin.eq}
\end{equation}
For a given wavevector $k_n$, buckling will become possible above a certain strain $e_n^*$ such that
$f_2^{3D}(e_{n}^*) = -Kk_n^2$. As $|f_2^{3D}|$ is an increasing function of the strain, the wavevector corresponding to the largest wavelength, i.e. the size of the box, will become unstable at the smallest strain, according to this analysis. This also implies that 
for a smaller box size, a larger strain would be needed to observe buckling; as noted above, in such small systems cavitation  tends to take place before the critical strain for buckling is reached, and the elastic analysis becomes irrelevant above the cavitation threshold.

In the following, numerical comparison between simulations and this theory will be
made by using for the elastic constants $C_{ij}$ values determined from simple linear deformations of a small sample that does not exhibit the buckling instability. These values are, for the interaction parameters and temperature mentioned below, $C_{11}=24.17$, $C_{33}=7.61$, $C_{23}=C_{13}=6.5$, and $G=0.07$ (in the same reduced units, the Young modulus of a glassy polymer is of order $50$, see section ``Materials and methods''). The Poisson ratio is $\nu = 0.178$.

\section{Size and strain rate dependence, interpretation}

The discussion in the previous section shows that the boundary conditions and geometry of the sample have an important influence on the buckling instability. For example, 
the instability takes place for a different mode for a system with free or with periodic boundary conditions, the first one undergoing a ``half wave'' instability which is prohibited in the second case. Also, the buckling of small samples is impossible as  the bending energy of the glassy phase is very high  compared to the deformation energy of the bulk. Therefore, a critical size  of sample can be defined as $L^*_y$. $L^*_y$, the minimal length from which the sample will be able to buckle under tensile strain before failure due to cavitation and crazing in the rubber phase becomes dominant. 
 
\begin{figure}
 \centering
\includegraphics[width=0.48\textwidth]{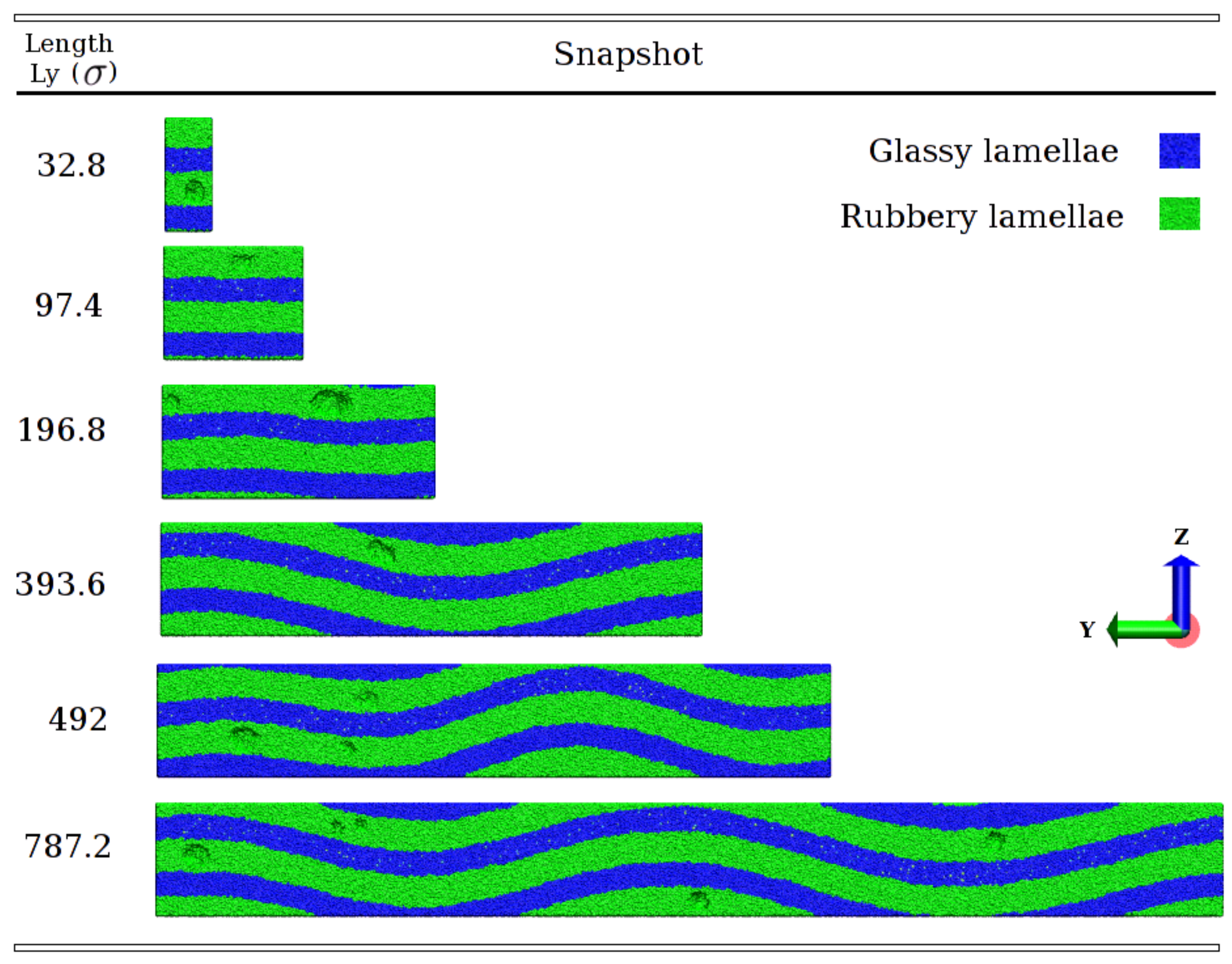}
\caption{Snapshots of several samples with different sizes. All configurations are taken at a true strain $e_{zz}=0.08$.  The samples are obtained by replication of a 
system with size $L_0=32.8\sigma$. Buckling is possible only  for the longer samples with $L_y\geq 393.6$). The buckling wave length seems to be size independent.}
\label{snapshot_sev_lenght.fig}
\end{figure}

 According to  elastic theory, the instability  will take place at smaller and smaller strains for bigger and bigger samples,
and always at the largest possible wavelength allowed by the boundary conditions. This, however, contradicts a number of experimental observations in which a rather well defined wavelength of the chevron structure is observed. This discrepancy is commonly assigned to preexisting defects in the microstructure \cite{Cohen_2_01}. Simulation provides an ideal benchmark of this hypothesis, as  it allows one to study  an ideal 
microstructure. We have therefore studied  samples of various sizes,  between $100$ and $800\sigma$ in the $Y$ direction. Figure \ref{snapshot_sev_lenght.fig}
 compares the  response of all tested samples, at a strain rate $7.3\times10^{-5}$. 
In terms of stress-strain relation all samples have roughly the same mechanical  behavior up to the yield point. 

However,  it is clear that  the minimal length for observing buckling before cavitation is 
 at least 200 monomer sizes $\sigma$, implying that such observations can be made only on very  large (by 
 simulation standards) systems.
 For samples larger than about  $400\sigma$, 
 the onset of  buckling occurs always  at the same strain ($e_{buck}=0.06$), in contradiction with the expectation from
 the elastic description of the previous section.  Another surprising observation, illustrated in figure \ref{snapshot_sev_lenght.fig}, is that the wavelength of the instability does not appear to increase with the size of the system. 

\begin{figure}[tp]	
\centering
\includegraphics[width=0.4\textwidth]{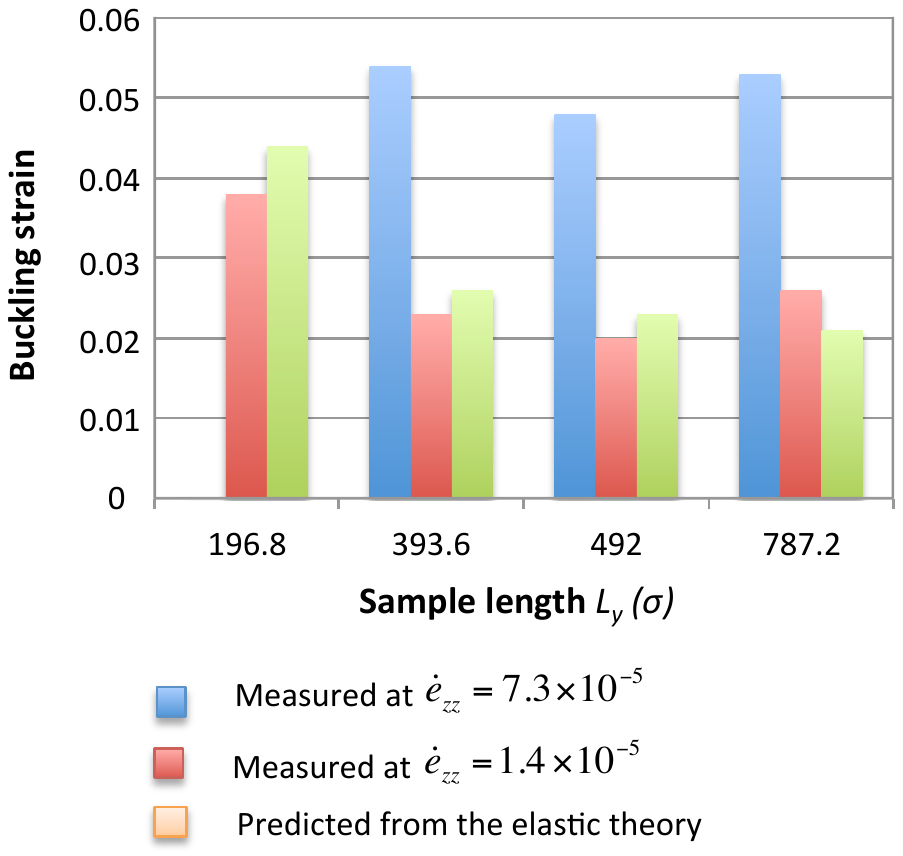}
 \caption{ Strain at buckling. Comparison for two different strain rates $7.3\times10^{-5}$ (blue bars) and $1.4\times10^{-5}$ (red bars)
for different system sizes, with the prediction of elastic theory (yellow bars).  Note that the smallest system does not exhibit buckling at high strain rate.
For large samples, the buckling at high and low strain rates are observed in different modes, see figures \ref{snap_buck.fig} and 
\ref{BucklingDeformationMap.fig}.
The comparison with the prediction of elastic theory
 shows that the latter is quite accurate when the buckling takes place in the 
largest wavelength mode, although it underestimates somewhat the buckling strain due to kinetic effects.} 
\label{barplot}
\end{figure}

Figure \ref{barplot} summarizes the buckling properties obtained at  strain rate of $7.3\times10^{-5}$. For each sample, the theoretical value of the buckling strain was calculated from equation \ref{buck_starin.eq}, assuming an instability wavevector $k=2\pi/L$ except for the largest sample.
The predicted values of $e_{zz}^{buck}$ is always less than the measured one. This difference will be interpreted below as  a direct consequence of kinetic factors that are  not taken into account in the elastic calculation.  Note, finally, that the  yield strain and stress are roughly the same for all samples. and are  correlated with the cavitation in the rubber phase.

\begin{figure}[tbp]
\centering
\includegraphics[width=0.5\textwidth]{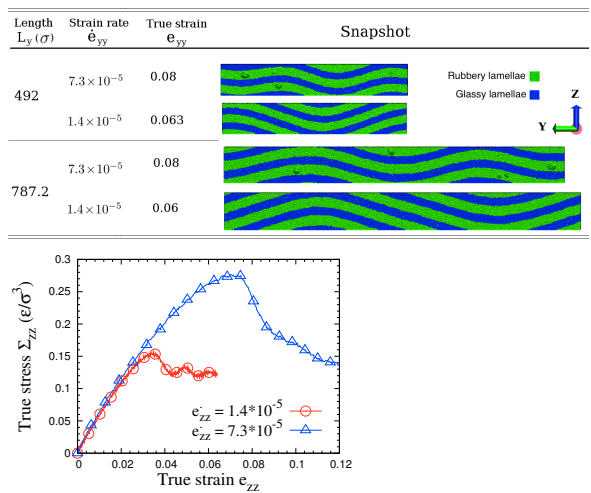}
\caption{ \label{snap_buck.fig}
Top panel: snapshots  of samples with   lateral size $L_y=492\sigma$  and $L_y=787\sigma$ driven at  two different strain rates ($\dot{e_{yy}}=1.4 \times 10^{-5}$ and $\dot{e_{yy}}=7.3 \times 10^{-5}$) . At low strain rate , the buckling wave length is equal to the sample length which is not the case at high strain rate. Note the imperfect aspect of the sinusoidal pattern in the first snapshot, indicative of a competition between modes with different wavelength. Note also that for this high driving rate, cavitation and buckling take place  simultaneously. 
Bottom panel: stress strain curves that correspond to the two different driving rates of the largest sample.  The high strain rate curve is identical to the one shown in figure \ref{behavior_curve.fig}
}
\end{figure}

In order to understand the role of strain rate, we have submitted the 
 same samples to a uniaxial tensile strain test driven at a lower strain rate ($\dot{\epsilon}_{yy} = 1.4\times10^{-5}$).
We find  that (i) the change in the Young modulus is negligibly small for all samples, (ii) the yield stress and strain decrease as the strain rate decreases and finally (iii) the stress softening exhibits a smooth transition (from yield to the drawing regime) at low strain rate compared to a large drop  at high strain rate.
Depending on sample size, the yield stress and strain are more or less affected. For the smallest sample, the decrease of the yield stress and strain is small compared to other samples.
 In general, the decrease of the yield threshold is strongly correlated with the change of the plastic mode from cavitation to buckling. Both cavitation and buckling result in a yield behavior, however the yielding associated with buckling is much more progressive and smooth than the one associated with cavitation, as illustrated in figure \ref{snap_buck.fig}. The most striking observation, however, is the fact that the wavelength of the buckling instability changes with strain rate. In the largest sample, it is equal to sample size, as expected from elasticity theory (see figure \ref{snap_buck.fig}).
 
\begin{figure}[tbp]
\includegraphics[width=0.48\textwidth]{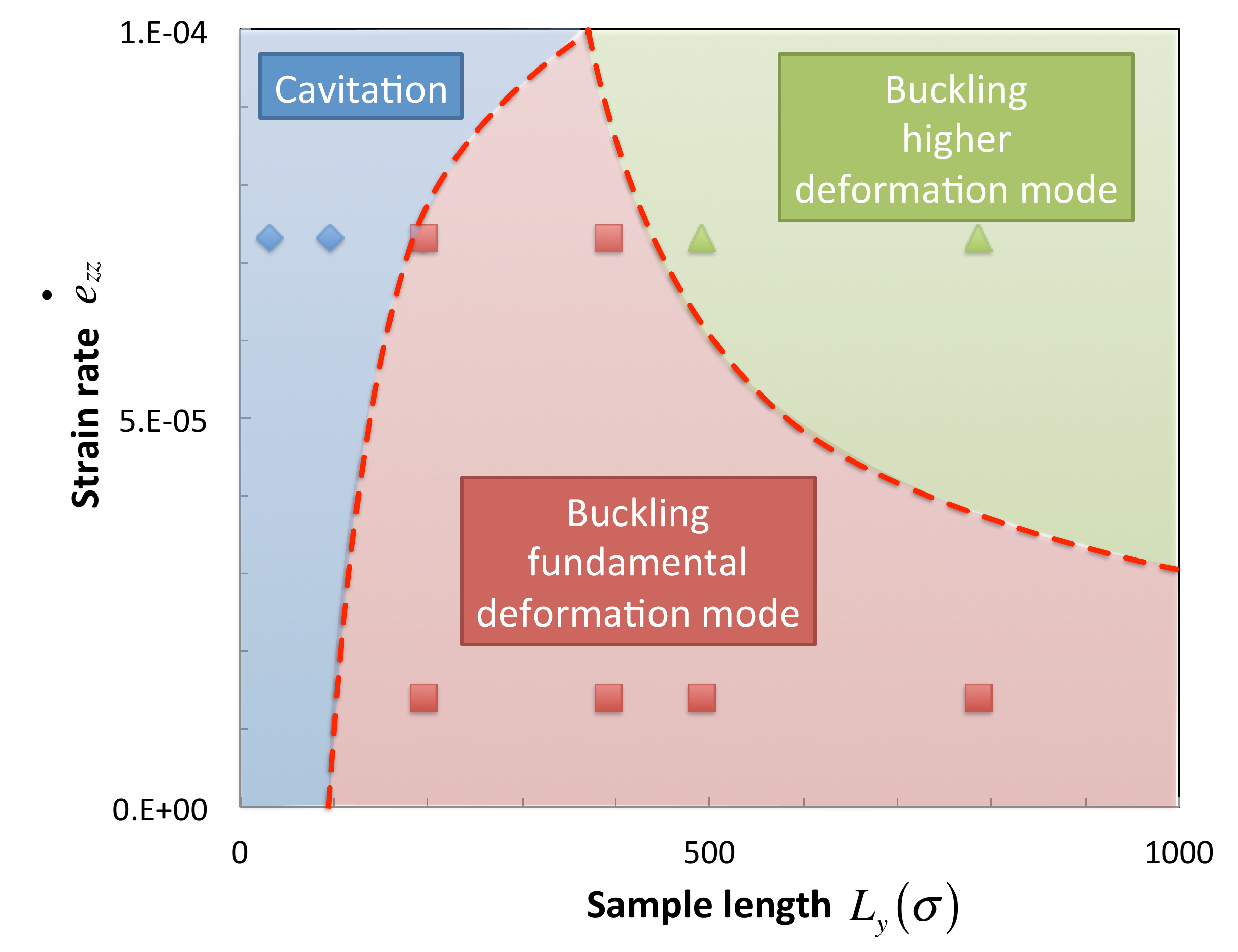}
\caption{\label{BucklingDeformationMap.fig}
Qualitative summary of our observations, indicating the various modes of deformation and yield depending on strain rate and sample size. The points indicate the simulated configurations, while the solid lines between regions are shown as guides to the eye.}
\end{figure}

 We propose an interpretation of this unexpected change of wavelength with strain rate based on a simple model of the competition between unstable modes.  
It has been argued above (section `` elastic description of the buckling instability'') that the onset of buckling at a certain wavevector $k$ is determined  by the function $F(e_{zz},k) = f_2(e_{zz})k_i^2+Kk_i^4 $
where $ f_2$ is obtained from equation \ref{f1_2D.eq} or \ref{f1_3D.eq}.   Above a critical strain $e_c$, $f_2(\epsilon)$ becomes negative,  and the wavevectors in the range 
$ 0\leq k \leq \sqrt{-f_2\/K}$ become linearly unstable.  We propose to account for the time evolution of the amplitude $U_k$ of a given mode trough the simple linear equation
\begin{equation}
\partial_t U_k =  - \Lambda F(e_{zz}(t),k) U_k
\label{eq:growth}
\end{equation}

where $\Lambda$ is a phenomenological kinetic coefficient which is assumed here, for simplicity, to be wavector and strain rate independent, at least in th erange of strain rates investigated in our simulations. Beyond $e_c$, the growth rate $\Lambda F$  has a maximum at a finite wavevector.  The key in understanding the effect of strain rate is in the time dependence of $F$, which is encoded in the time dependence of $e_{zz}$ and renders  the evolution nontrivial.  For small strain rates, the instability of the  smallest wavevector $k=2\pi/L$ has ample time to develop before the strain increases and makes a  second mode at $k=4\pi/L$ unstable. At higher strain rates, the growth rate evolves on time scales comparable to the instability itself, and a competition between modes arises. By solving numerically equation \ref{eq:growth} with a value of $\Lambda$ adjusted to reproduce the observed growth of the instability in one of the configurations, we found that, for the largest sample size considered in our
simulations, the mode that corresponds to $k=4\pi/L$ indeed overwhelms the long wavelength mode $k=2\pi/L$ when the system is driven at the largest strain rate, before the latter mode has time to develop significantly.  Therefore the second mode, rather than the first, is observed. While the very high strain rates used in simulation (see section ``Material and method'') result here in a selection of very short wavelengths of the order of nanometers, the effect is expected to be very generic, and could result in the submicron chevron structures observed in experiments. These results are summarized in figure \ref{BucklingDeformationMap.fig}, which displays the  competing mechanisms depending on sample size and strain rate.

\section{Conclusion}
\label{sec:conclusion}

The buckling instability of layered materials has been explored here for the first time by atomistic simulations, in the case on a lamellar copolymer
with glassy and rubbery lamellae. This approach allows one to identify unambiguously
the mechanism that drives the instability and failure of such materials, as being caused by the elastic contrast between the two phases and a Poisson ratio effect. 
Still, a pure elastic theory is insufficient to account for the dependence of the observed instability on strain rate, which is also reported here for the first time. 
A simple description of this dependence was provided by a model that involves the competition between the growth rate of linearly unstable modes with the  rate
of deformation. This rate dependence of the deformation mode of the microstructure, which is still to be investigated in controlled experiments, indicates the possibility of  a rich and still unexplored phenomenology in the mechanical behavior of layered materials. The generality of the mechanism for buckling points to  the importance of this 
instability as a new  plasticity mechanism in nanostructured polymer materials, which is expected to operate in semicrystalline materials as well as in block copolymers, and to  complement  more traditional mechanisms such as cavitation and crazing.

\section{Materials and methods}
 \label{sec:methods}
Our molecular dynamics simulations involve a standard coarse grained model  polymer chains.  Each chain is represented by 
a sequence of 50-100-50  beads of different chemical nature ($A$ and$B$) interacting via covalent FENE bonds and Lennard
Jones interparticle potentials. The parameters of the LJ potentials are chosen to ensure that phase separation into a lamellar morphology takes place,
and that the $A$ layers are in a glassy state while the $B$ layers remain rubbery at the tensile test temperature ($T_{test}\simeq300K$).  The method used for generating samples has been detailed in 
reference \cite{Perez08}. In terms of Lennard-Jones interaction parameters, we take 
$\epsilon_{AA}=1$,  $\epsilon_{BB}=0.3$, $\epsilon_{AB}=0.4$. 
The resulting glass transition temperatures  are $T_g^A\simeq420K$ for the $A$ layers and $T_g^B\simeq200K$ for the $B$ layers.
The diameter and mass, $\sigma$  and $m$, are
identical for all species, and serve as length and mass unit, respectively.  Using an energy scale of $1000K/k_B$ and a length scale of 
 $0.5 nm$ which are typical in the coarse grained descriptions of standard polymers, the corresponding stress 
unit is of order 100MPa, and the Young modulus of the glassy polymer is of the order of 1-10GPa.

An important remark here is that the time scale that results from these choices of units, if parameters appropriate for typical polymers
are used, lies in the picosecond time range. Therefore the strain rates achieved in simulations are of the order of $10^7 s^{-1}$ in real units, extremely high compared to typical experimental rates. As is often the case in simulation studies involving glassy materials, the behavior observed in simulation studies must be understood as being qualitatively, rather than quantitatively, representative of the experimental reality. 

{\bf Acknowledgments}

 Computational support by the F\'ed\'eration Lyonnaise de Calcul Haute Performance and Grand Equipement National de Calcul Intensif  GENCI-CINES are acknowledged. Financial support from ANR   (Nanomeca project)
 and R\'egion Rh\^one-Alpes (Macodev project) are also
  acknowledged. JLB is supported by Institut Universitaire de France.
  Part of the simulations were carried out using  the LAMMPS molecular dynamics software
 (http://lammps.sandia.gov). We thank Prof. D.J. Read and Prof. T. McLeish (Leeds University) for useful exchanges.

%

\end{document}